\documentclass[13pt]{amsart}

\usepackage{amsmath,bbold}

\usepackage{times,graphicx}
\usepackage{bm}
\usepackage{natbib}

\calclayout

\begin{document}
\begin{center} {\Large An illustration of the risk of borrowing information via a shared likelihood} \\

{P. Richard Hahn}\\
Arizona State University
\end{center}
\vspace{0.3in}
\begin{center} Abstract \end{center}
A concrete, stylized example illustrates that inferences may be degraded, rather than improved, by incorporating supplementary data via a joint likelihood. In the example, the likelihood is assumed to be correctly specified, as is the prior over the parameter of interest; all that is necessary for the joint modeling approach to suffer is misspecification of the prior over a nuisance parameter. 

\section{Borrowing information via a joint likelihood}
Suppose we observe a single draw from $Y \sim \mbox{N}(\theta, 1)$ and that interest lies in estimating $\mbox{E}(Y) = \theta$ for $\theta \in \lbrace 0, 1 \rbrace $. Suppose further that we also observes a single, independent draw from $X  \sim \mbox{N}(\theta + \mu, 1)$ with $\mu$ unknown.  Finally, fix $\mbox{P}(\theta = 0) = \mbox{P}(\theta = 1) = \frac{1}{2}$. Our estimand is $\theta$, while $\mu$ is a nuisance parameter. We will evaluate estimators $\hat{\theta}$ according to 0-1 classification loss 
\begin{equation}
L(\theta, \hat{\theta})  := \mathbb{1}\lbrace \hat{\theta} \neq \theta \rbrace,
\end{equation}
for which the posterior mode is the associated Bayes estimator. In particular, under our uniform prior over $\theta$, the Bayes estimator simply sets $\hat{\theta}$ to whichever value of $\theta$ yields the highest likelihood evaluation. The question this paper examines is when $\hat{\theta}_{xy} := \mbox{arg max}_{\theta} f(x,y \mid \theta)$ should be preferred over $\hat{\theta}_{y} := \mbox{arg max}_{\theta} f(y \mid \theta)$.  

Consider the risk $R(\hat{\theta}) := \mbox{E}_{X, Y, \theta}\lbrace L(\theta, \hat{\theta}) \rbrace$. Specifically, $R(\hat{\theta}_{y})$ and $R(\hat{\theta}_{xy})$ have convenient expressions,
\begin{equation}
\begin{split}
R(\hat{\theta}_{y}) =& \frac{1}{2} \mbox{Pr}_{\theta = 1}\lbrace\phi(Y) > \phi(Y-1)\rbrace + \frac{1}{2} \mbox{Pr}_{\theta = 0}\lbrace\phi(Y) < \phi(Y-1)\rbrace,
\end{split}
\end{equation}
and
\begin{equation}
\begin{split}
R(\hat{\theta}_{xy}) =& \frac{1}{2} \mbox{Pr}_{\theta = 1}\lbrace\phi(Y)\phi(X-\mu) > \phi(Y-1)\phi((X-1-\mu))\rbrace + \\ 
&\frac{1}{2} \mbox{Pr}_{\theta = 0}\lbrace\phi(Y)\phi(X-\mu) < \phi(Y-1)\phi(X-1-\mu)\rbrace,
\end{split}
\end{equation}
where $\phi(\cdot)$ denotes the standard normal probability density function.

Working inside the probability braces, we may take logarithms on both sides of the inequality and simplify algebraically. The first expression simply becomes $R(\hat{\theta}_{y}) = \Phi(-1/2)$, where $\Phi(\cdot)$ denotes the standard normal cumulative distribution function. The expression for $R(\hat{\theta}_{xy})$ will involve the unknown nuisance parameter $\mu$. Taking a Bayesian approach, we integrate over this parameter, using distribution $\mbox{N}(0, w^2)$, which implies that $X \sim \mbox{N}(0, 1 + w^2)$. For notational simplicity, we will write $s^2 := 1 + w^2$. Making this substitution we have
\begin{equation}\label{joint_risk}
\begin{split}
R(\hat{\theta}_{xy}) =& \frac{1}{2} \mbox{Pr}_{\theta = 1}\lbrace\phi(Y)\phi(X/\sigma) > \phi(Y-1)\phi((X-1)/s)\rbrace + \\ 
&\frac{1}{2} \mbox{Pr}_{\theta = 0}\lbrace\phi(Y)\phi(X/\sigma) < \phi(Y-1)\phi((X-1)/s)\rbrace,\\
 =&\frac{1}{2} \mbox{Pr}_{\theta = 1}\lbrace Y < -X/s^2 + (1/s^2 + 1)/2 \rbrace +\\
&\frac{1}{2} \mbox{Pr}_{\theta = 0}\lbrace Y > -X/s^2 + (1/s^2 + 1)/2 \rbrace.
\end{split}
\end{equation}
It is easy to see that as $s \rightarrow \infty$, $R(\hat{\theta}_{xy}) \rightarrow R(\hat{\theta}_{y})$.  

\section{Nuisance parameter prior misspecification}
Though expression (\ref{joint_risk}) obscures the fact, the outer probability evaluation depends on the true distribution of $X$. That is, assuming the rest of the model is correctly specified, the risk performance of $\hat{\theta}_{xy}$ still depends on proper selection of the nuisance hyper-parameter $s$. To foreground this fact, for $\sigma^2 \geq 1$, define $\sigma^2 - 1$ to be the variance of nature's distribution over the nuisance parameter $\mu$, so that $X \sim \mbox{N}(0, \sigma^2)$. What is the effect of (mis)specifying $s^2 \neq \sigma^2$?

\begin{figure}
\begin{center}
\includegraphics[width=4in]{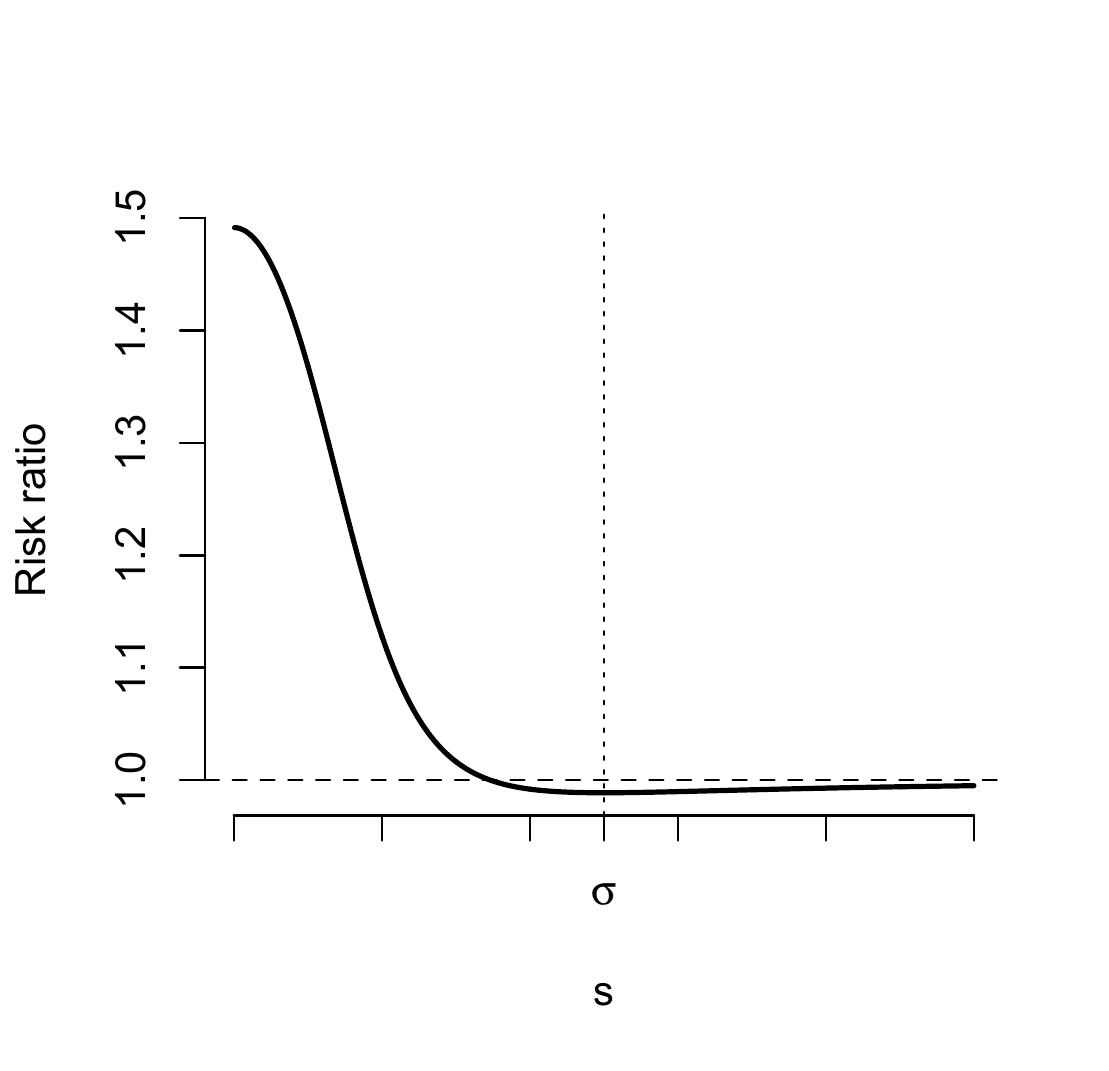}
\caption{For $s < \sigma$, the risk ratio can be substantially greater than one. For $s \gg \sigma$, the risk ratio approaches one. The challenge in practice is that $\sigma$ is not known in advance, nor is it learnable from the data. Moreover, the improvement that comes with modeling $X$ with the appropriate $s = \sigma$ is modest relative to the cost of under-specifying $s$. }\label{risk_ratio}
\end{center}
\end{figure}

Write $a = -(1/s^2 + 1)/2$ and $b = 1/s^2$ and define $Z := Y + bX + a$. Next, observe that $\mbox{E}(Z \mid \theta = 0) = a$,  $\mbox{E}(Z \mid \theta = 1) = 1 + b + a$ and $\mbox{Var}(Z \mid \theta = 0) = \mbox{Var}(Z \mid \theta = 1) = 1 + b^2\sigma^2$, from which it follows that
\begin{equation}\label{joint_risk2}
R(\hat{\theta}_{xy}) = \frac{1}{2} \left \lbrace 1 - \Phi \left(\frac{-a}{\sqrt{1 + b^2\sigma^2} }\right) \right \rbrace + \frac{1}{2}\Phi \left(\frac{-1-b-a}{\sqrt{1 + b^2\sigma^2}}\right).
\end{equation}
This formulation allows plotting of the risk ratio $R(\hat{\theta}_{xy})/R(\hat{\theta}_{y})$ as a function of the hyper-parameter choice $s$. Figure \ref{risk_ratio} shows that when $s$ is chosen approximately correctly, $s \approx \sigma$, one realizes the anticipated improvement from ``borrowing information" from $x$.  However, if $s$ is chosen too small, the risk performance of $R(\hat{\theta}_{xy})$ is worse than if one had just ignored $x$ entirely. What is more, the potential risk amplification is seen to be much more severe than the possible risk reduction.

\section{Discussion}
Model based inference using multiple data sources is often advertised as an unequivocal virtue \citep{unlabeled}. At the time of writing, an internet search of scholarly papers from the past ten years that contain the phrases ``borrowing information" and/or ``joint model" turns up thousands of results.  

Incorporating ``side data'' requires making additional modeling assumptions, which present fresh opportunities for model misspecification. The example above demonstrates that an estimator based on a misspecified joint model can have markedly worse risk performance than a correctly specified model that ignores some elements of the data entirely.  Moreover, the example shows that the degraded risk performance occurs even under seemingly superficial misspecification --- the likelihood is correct, the prior over the parameter of interest is correct, it is only the prior over a nuisance parameter that is faulty.  

Though stylized, this example retains key features of modern applied Bayesian models. In particular, although the sample size of one is itself unrealistic, the root of the difficulty is that $\sigma$ is unidentified by the available data, which puts the onus of good risk performance on the judicious selection of prior; truly this is a feature of most modern Bayesian nonparametrics (although there is more data, there are even more parameters). Moreover, some real-world problems have a structure very similar to the one here --- consider the case where a small, carefully collected sample is augmented with a much larger amount of crowd-sourced data, which tends to be of much lower fidelity.

One could argue that vague priors seem to do no harm; in the limit it is as if the side data were being ignored.  But this position leaves aside two important points.  First, one never knows for sure what is vague enough. Second, it ignores the often substantial effort required to model the side data in the first place.


Applied statisticians undertaking the construction of elaborate joint models would be well advised to consider carefully if the problem at hand is, like the example here, a situation where more (data) may be less (statistically reliable).
\bibliographystyle{abbrvnat}
\bibliography{misborrow}

\end{document}